\def\ts     {\thinspace}
\def\kms    {\ifmmode{{\rm \ts km\ts s}^{-1}}\else{\ts km\ts s$^{-1}$}\fi}
\def\mo     {M$_{\odot}$}
\def\hi     {H{\small I}}

\documentclass{aa}
\usepackage{graphics} 
\begin{document}
  \thesaurus{ 03(09.13.2, 11.04.2, 11.06.1, 11.09.1 NGC\,3077, 
   11.09.4, 13.19.3) }  
  \title{ Extensive molecular gas in the tidal arms near NGC\,3077 --
Birth of a dwarf galaxy? }
  \author{Andreas Heithausen\inst{1}
\and
Fabian Walter\inst{2}}
  \institute{ 
Radioastronomisches Institut der Universit\"at Bonn, Auf dem H\"ugel 71,
D-53121 Bonn, Germany, E-mail: heith@astro.uni-bonn.de
\and
California Institute of Technology,
  Astronomy Department 105-24, Pasadena, CA 91125,
  U.S.A., Email: fw@astro.caltech.edu}

  \offprints { A. Heithausen}
  \date {Received 28 April 2000, Accepted 21 June 2000}
\authorrunning{Heithausen \& Walter}
\titlerunning{Molecular emission near NGC\,3077}

\maketitle

\begin{abstract}
Using the IRAM 30\,m radio telescope we have mapped the tidal arm
feature south--east of NGC\,3077 where we recently detected molecular
gas in the CO ($J$=1$\to$0) and (2$\to$1) transitions. We find that
the molecular gas is much more extended than previously thought
(several kpc).  The CO emission can be separated into at least 3
distinct complexes with equivalent radii between 250\,pc and 700\,pc
and all well confined over a narrow range in velocity
-- the newly detected complexes therefore range among the largest
molecular complexes in the local universe.  For one complex we have
also obtained a CO (3$\to$2) spectrum using the KOSMA 3\,m radio
telescope; utilizing an LVG model we find that the kinetic temperature
for this complex must be about 10\,K, and the H$_2$ volume density
between 600 and 10000\,cm$^{-3}$. Mass estimates based on
virialization 
yield a total mass
for the complexes of order $2-4\times10^7$\,\mo, i.e. more than the
estimated molecular mass within NGC\,3077 itself. This implies that
interactions between galaxies can efficiently remove heavy elements
and molecules from a galaxy and enrich the intergalactic medium.  A
comparison of the distribution of \hi\ and CO shows no clear
correlation. However, CO is only found in regions where the \hi\
column density exceeds $1.1\times10^{21}$\,cm$^{-2}$. \hi\ masses for
the molecular complexes mapped are of the same order as the
corresponding molecular masses. 
Because the intergalactic pressure is most likely too low
to confine the complexes we conclude that they are gravitationally bound.
Since the tidal arm with its molecular complexes
 has all the ingredients to form
stars in the future, we are thus presumably witnessing the birth of a
dwarf galaxy. 
This process could be important for the formation of dwarf galaxies
especially at larger look--back times in the universe where
galaxy interactions may have been more frequent.

 \end{abstract}

\section{Introduction}

Over the last decades it has become clear that tidal forces during
close encounters of galaxies can redistribute large masses as long
tails or bridges between the interacting galaxies.  The idea that part
of these newly formed structures could form self--gravitating entities
was proposed by Zwicky (\cite{zwicky56}). Numerical simulations
(e.g. Elmegreen et al.\ \cite{elmegreen:etal93}; Barnes \& Hernquist
\cite{barnes:hernquist96}) support this scenario.

Observationally, tidal systems are best traced by the neutral gas
phase (by means of \hi\ observations) since it is this extended
component of a galaxy which is most easily disrupted by interactions.
Also, much work has been done in studying the tidal arms of
interacting galaxies at optical wavelengths to find regions of active
star-formation (`tidal dwarfs').  Impressive examples of tidal arms
with on-going star-formation are, e.g., the Antenna galaxy
(NGC\,4038/39, Mirabel et al. \cite{mirabel:etal92}) and the
Superantenna system (Mirabel et al.\ \cite{mirabel:etal91}).  Based on
an optical study of 42 Hickson compact groups of galaxies Hunsberger
et al.\ (\cite{hunsberger:etal96}) speculate that up to 50\% of the
dwarf galaxies in such compact groups might be created by tidal
interaction among giant parent galaxies.

\begin{figure*}
\rotatebox{-90}{
 \resizebox{7cm}{!}{\includegraphics{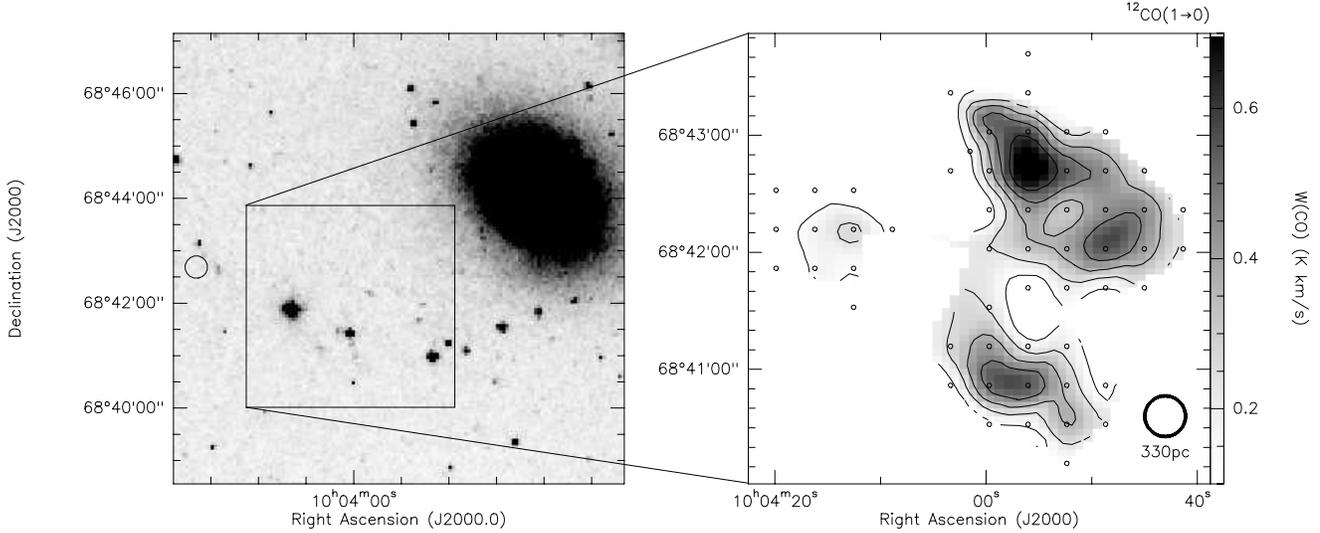}}}
\caption{
Overview of the observed region. {\em Left:} optical image of
NGC\,3077 (elliptical object to the north--west) and its surrounding
(taken from the digitized sky survey). For a complete optical picture
of the M\,81 triplet see Walter \& Heithausen
(\cite{walter:heithausen99}).  The box indicates the region which
we mapped in CO; 
the open circle indicates the position towards which we
searched for CO but couldn't detect any. {\em Right:} Blowup of our
integrated CO map (J=1$\to$0 transition). Contours are every
0.12\,K\,\kms\ ($2\sigma$) starting at 0.12\, K\,\kms. Observed
positions are indicated as small circles. The beamsize is indicated in
the lower right corner of the map; it corresponds to 330\,pc at an
adopted distance of 3.2\,Mpc.}
\label{comap}
\end{figure*}

\begin{table*}
\caption{Observed and derived parameters for molecular complexes}
\begin{tabular}{l l l l l l l l l l}
\noalign{\hrule}
\noalign{\smallskip}
\# & $\alpha_{J2000}$ & $\delta_{J2000}$ & radius & $W$(CO) & $v_{hel}$ & 
$\Delta v$ & $M_{vir}$ & $M_X^{(a)}$ & $<N($\hi$)>$\\
   &                  &         & (pc) & (K\,\kms)&(\kms)    & (\kms)  
& (\mo) & (\mo) & $10^{21}$\,cm$^{-2}$ \\   
 \noalign{\hrule\smallskip}
1 & $10^h03^m56.\!^s0$ & $68^\circ42' 41.\!''8$ & 700      & 
$0.45\pm0.02$ & $13.9\pm0.3$ & $13.5\pm0.8$ & 2.4\,10$^7$ & 1.7\,10$^7$ & 1.6\\
2 & $10^h03^m56.\!^s0$ & $68^\circ40' 51.\!''8$ & 390      &
$0.34\pm0.02$ & $17.2\pm0.4$ & $14.7\pm1.0$ & 1.6\,10$^7$ & 0.4\,10$^7$ & 1.6\\
3 & $10^h04^m16.\!^s5$ & $68^\circ42' 11.\!''8$ & $\ge250$ & $0.21\pm0.02$ &
$20.4\pm0.4$ & $5.2\pm0.6$ & $-$ & $\ge0.1\,10^7$ & 1.4\\
4 & $10^h04^m32.\!^s8$ & $68^\circ42' 41.\!''8$ & $-$ & $\le0.06$ &
 $-$ & $-$ & $-$ & $-$ & 1.1 \\
 \noalign{\hrule\smallskip}
\noalign{\noindent Remarks: Adopted distance 3.2\,Mpc; $radius=\sqrt{area/\pi}$;
line parameters are derived from a gaussian fit to the cloud averaged spectrum.
a: $X_{\rm CO}=8\,10^{20}$\,cm$^{-2}$\,(K\,\kms)$^{-1}$\ adopted}
\label{coclouds}
\end{tabular}
\end{table*}

Surprisingly little is known about molecular gas in tidal arms around
interacting galaxies.  However, this is an important issue since
molecular clouds are the places where stars are born. The distribution
of molecular gas in quiescent extragalactic objects (such as tidal
arms) therefore gives clues as where to expect star formation to
commence in the future.

Combes et al. (\cite{combes:etal88}) were the first to detect a
CO cloud outside the optical body of a galaxy
(NGC\,4438). Another small molecular cloud in a tidally influenced
environment was detected by Brouillet et al.\
(\cite{brouillet:etal92}) in a torn--out spiral arm of M\,81. Smith et
al.\ (\cite{smith:etal99}) discovered a molecule rich tail in Arp\,215
and argue that its metal rich gas has been driven out from the inner
disk of the parent galaxy. CO--emission has also been detected in the
outskirts of Cen\,A which can be probably attributed to its merger
history (Charmandaris et al.\ \cite{charmandaris:etal2000}). Braine et
al. (\cite{braine:etal2000}) described two further molecular clouds
associated tidal arms of interacting galaxies, where star-formation
already is taking place.

The most extended molecular complex in tidal arms of interacting
galaxies was recently discovered by us near NGC\,3077 (Walter \&
Heithausen \cite{walter:heithausen99}), member of the M\,81
triplet. This complex is of particular interest since, although it is
huge, hardly any star formation seems to be associated with it.

In this paper we present a detailed follow-up study of this complex
(Walter \& Heithausen
\cite{walter:heithausen99}). Sect.~\ref{observations} briefly describes
our new CO observations obtained with the IRAM 30\,m and the KOSMA
3\,m radio telescopes. Our results are presented in
Sect.~\ref{results}. In Sect.~\ref{excitation} we analyze the excitation
condition of the CO gas using our multi--transition observations and a
simple LVG--model to match the observed line ratios. Masses for the
molecular complexes are estimated in Sect.~\ref{comass}. A critical
parameter for molecular gas is the shielding column density against
the destroying UV radiation; we analyze the relation between \hi\ and
CO in the tidal arms in Sect.~\ref{HI}. The origin and future of the
complexes are discussed in Sect. \ref{origin}.  We summarize our
results and conclusions in Sect.~\ref{conclusions}, discuss the
implications for metal loss and the enrichment of the intergalactic
medium (IGM) due to gravitational interactions and speculate that we
may be witnessing the birth of a new dwarf galaxy.

\section{Observations \label{observations}}

The CO ($J$=1$\to$0) and (2$\to$1) transitions have been observed in
July 1999 using the IRAM 30\,m radio telescope. The beam sizes are
22$''$ at 115\,GHz and 11$''$ at 230\,GHz (corresponding to 330\,pc
and 165\,pc at the adopted distance of NGC\,3077, 3.2\,Mpc). The
main beam efficiencies are $\eta_{\rm mb}(115 \rm GHz)=0.84$ and
$\eta_{\rm mb}(230 \rm GHz)=0.55$, respectively.  Pointing accuracy
was better than 5$''$. Spectra were obtained with a velocity
resolution of 0.8\,\kms\ at 115\,GHz and 0.4\,\kms\ using
autocorrelator spectrometers. In total, we observed 57 individual
positions simultaneously in both transitions with a wobbling secondary
mirror; wobbler throw was $\pm4'$ in azimuth.  The spacing between
individual positions is 20$''$ ($\sim$ the size of the beam at
115\,GHz).

In January 2000 we also obtained one CO spectrum in the ($J=3\to2$)
transition using the KOSMA 3\,m radio telescope located at the
Gornergrat near Zermatt in the Swiss Alps. This spectrum was obtained
with a wobbling secondary mirror with a throw of $\pm3'$ in azimuthal
direction. We used a medium resolution acusto-optical spectrometer
with a velocity resolution which finally has been degraded to
2.3\,\kms. The main beam efficiency is $\eta_{\rm mb}=0.70$, the beam
size 80$''$ (FWHM). The final rms is 4.8mK ($T_{\rm mb}$).

\section{Results \label{results}}


Fig.~\ref{comap} gives an overview over our mapping results. An
optical image is shown on the left (as obtained from the Digitized Sky
Survey, DSS).  Our CO detections (as indicated by the box and the `X')
are clearly located outside NGC\,3077 (the elliptical object
north--west of the centre). The area where we obtained an almost
complete map in the 2 lowest CO transitions is shown in
Fig.~\ref{comap} (right).  The CO emission is clearly extended over
several kpc (1$'\sim$1\,kpc) and can be subdivided into at least two
separate complexes. A further cloud (`X') has been detected outside
the mapped area. This area has only partly been mapped by us as yet
and an average spectrum is presented in Fig.~\ref{fig_cloud3}.

\begin{figure}
 \resizebox{8.6cm}{!}{\includegraphics{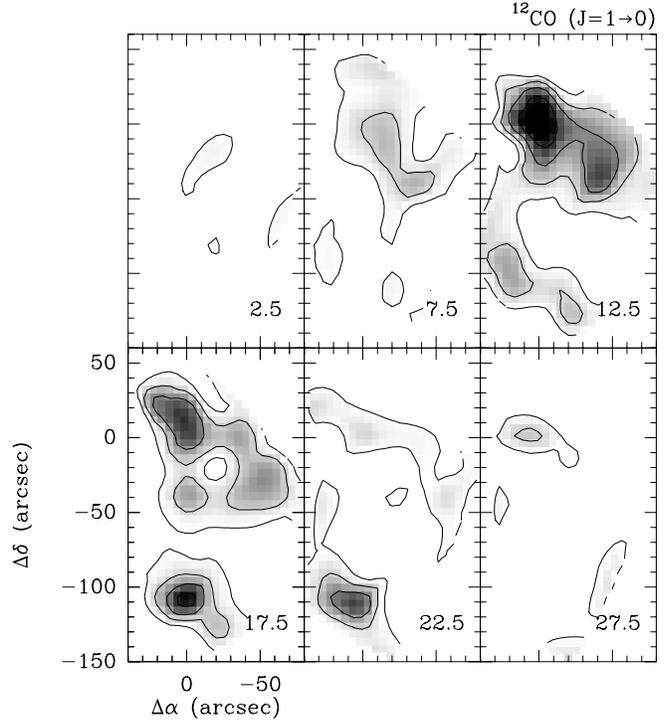}}
\caption{
Channel maps of complexes \#1 and 2, averaged over 5\,\kms. Center
velocities of each channel are indicated in the lower right
corner. Contours are every 0.012 K ($2\sigma$) starting at
0.012\,K. Offsets are relative to $\alpha_{J2000}=10^h03^m56.\!^s0$;
$\delta_{J2000}=68^\circ42' 41.\!''8$ }
\label{cochan}
\end{figure}

\begin{figure}
\rotatebox{0}{
 \resizebox{8.6cm}{!}{\includegraphics{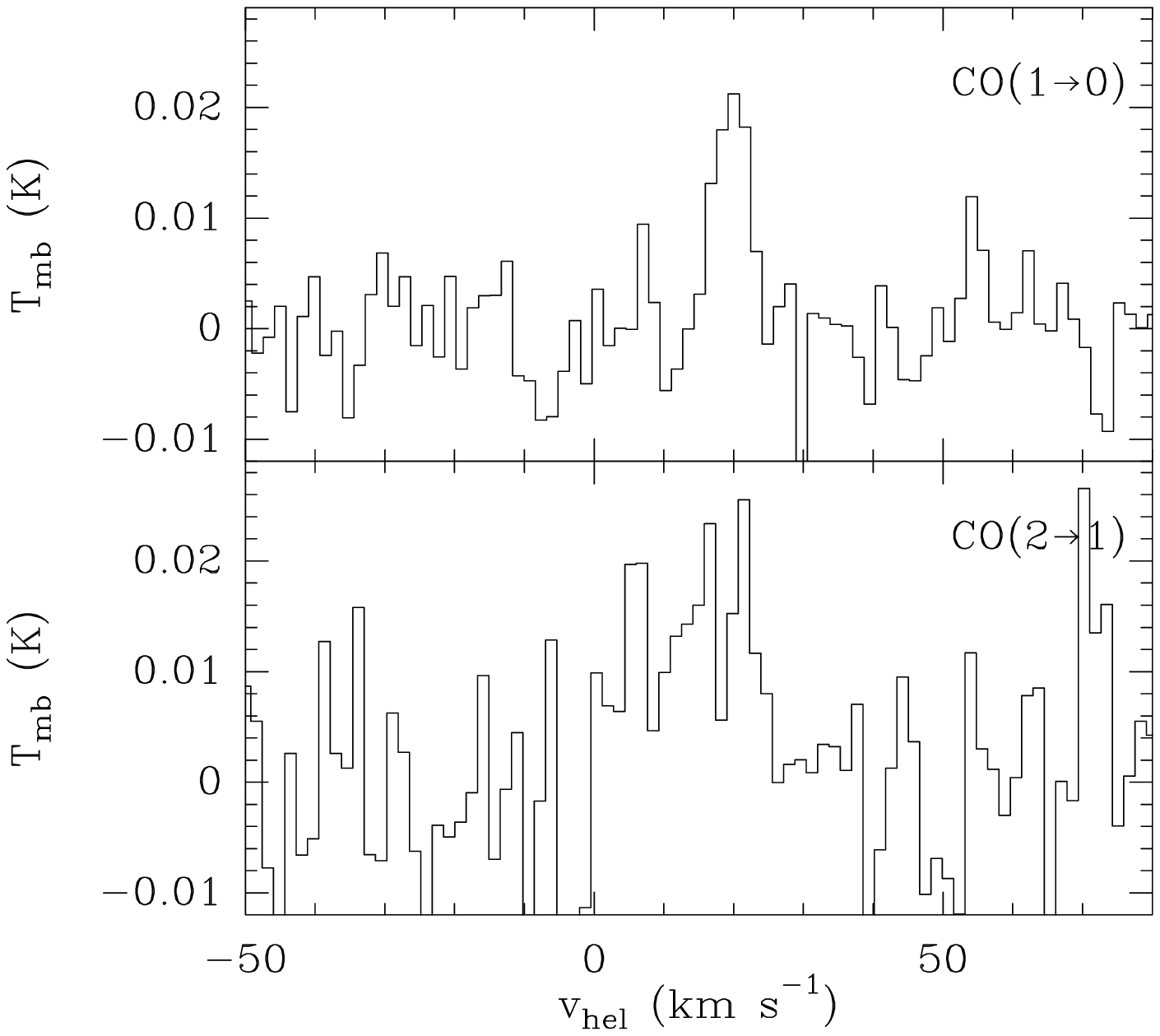}}}
\caption{Average CO  $(J=1\to0)$ (top)and  $(J=2\to1)$ (bottom)
spectra of complex \#3. }
\label{fig_cloud3}
\end{figure}

Table \ref{coclouds} summarizes some observed and derived parameters
for the three detected complexes. It also list a fourth position in the
direction of a `Garland' object (Karachentsev et
al. \cite{karachentsev:etal85a}) where we could not detect CO 
(marked as a open circle in Fig.~\ref{comap}, left). In
Table~\ref{coclouds} the integrated CO $(J=1\to0)$ line intensity,
$W$(CO), the center velocity, $v_{\rm hel}$, and the full width at
half maximum of the lines, $\Delta v$ are derived from spectra
averaged over the area with significant line emission of the single
complexes. The radius is an equivalent radius of a circle surrounding the
same area as where CO emission was detected
($radius=\sqrt{area/\pi}$).

The velocity structure of complexes \#1 and \#2 is visible in the channel
maps displayed in Fig.~\ref{cochan}. Cloud \#1 is extended from the
south--west to north--east. Complex \#2 is the southern more compact
object. It shows evidence for further substructure.
We expect these complexes to break
up in more little clouds when observed at higher spatial resolution
(see also the discussion in Sect.~\ref{origin}). Throughout the paper
we discuss only properties of the whole complexes and do not subdivide
them further, although especially for complex \#1 there is evidence
for substructure in both the channel maps 
(Fig.~\ref{cochan}) and in the integrated map (Fig.~\ref{comap}).

\begin{table}
\caption{Line parameter of CO spectra of complex \#1}
\begin{tabular}{l l l l l}
\noalign{\hrule}
\noalign{\smallskip}
Transition & $T_{mb}$ & $rms$  & $v_{hel}$ & $\Delta v$ \\
           & (mK)      & (mK)    & (\kms)    &   (\kms) \\
\noalign{\smallskip\hrule\smallskip}
$(1\to0)$ & 29 & 2 & $13.9\pm0.3$ & $13.4\pm0.7$ \\
$(2\to1)$ & 17 & 2 & $11.4\pm0.7$ & $15.0\pm1.6$ \\
$(3\to2)$ & $-$ & 4.8 & $-$ & $-$ \\
\noalign{\hrule\smallskip Remarks:
The (1$\to$0) and (2$\to$1) spectra
have been convolved to an angular resolution  of 80$''$, the size of the KOSMA
beam at 345\,GHz.}
\label{tab_cloud1}
\end{tabular}
\end{table}

\begin{figure}
\resizebox{8cm}{!}
{\includegraphics{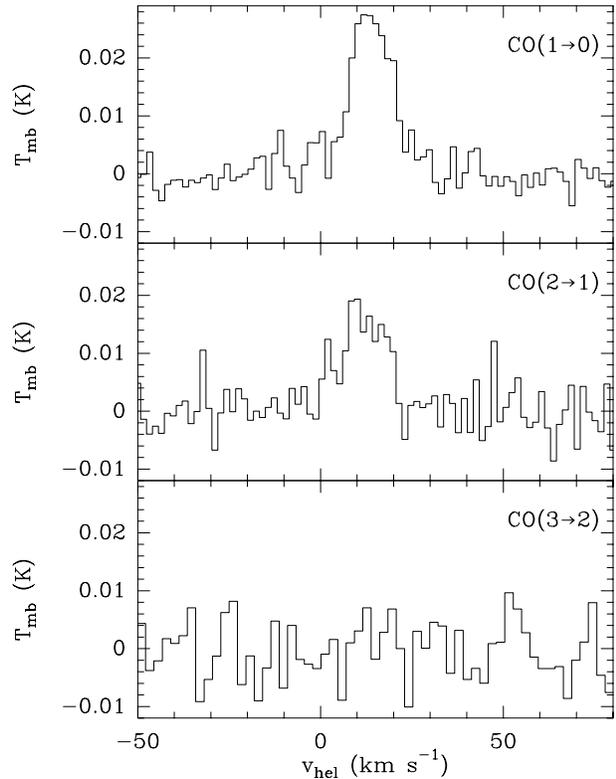}}
\caption{CO spectra towards complex \#1. The $(1\to0$) and (2$\to$1) spectra
have been convolved to an angular resolution of 80$''$ (1.3\,kpc), the
size of the KOSMA beam at 345\,GHz.}
\label{ph321}
\end{figure}

In order to constrain the excitation conditions of the more extended,
northern complex, we have observed it with the KOSMA 3\,m radio
telescope in the CO (3$\to$2) transition. The spectrum is shown in
Fig. \ref{ph321} together with the (1$\to$0) and (2$\to$1) transitions
convolved to the same angular resolution as that of the KOSMA spectrum
(80$''$, 1.3\,kpc). The corresponding values derived from a gaussian
analysis of the spectra are listed in Table~\ref{tab_cloud1}. No CO
(3$\to$2) line emission was detected. The integrated line ratio for
the lower two CO transitions is $R_{21}=W(2\to1)/W(1\to0)=0.66$ with
no significant variation throughout complex \#1. The upper limit for the
ratio of the (3$\to$2) line to the (2$\to$1) line is
$R_{32}=W(3\to2)/W(2\to1)\le0.1$; implications on the excitation
conditions are discussed in Sect.~\ref{excitation}.

\begin{figure*}
\rotatebox{-90}{
 \resizebox{8.5cm}{!}{\includegraphics{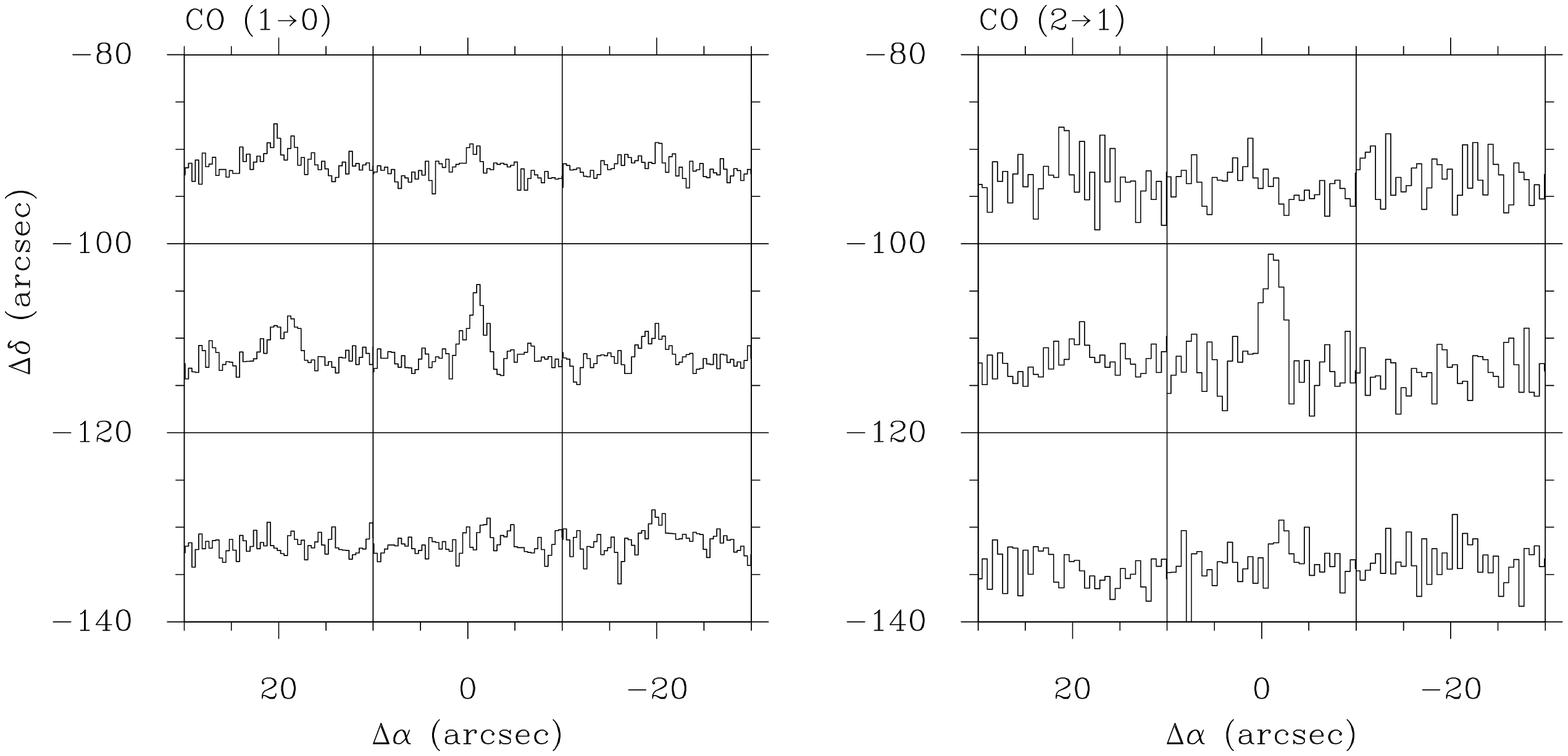}}}
\caption{
CO ($J=1\to0$) (left) and (2$\to$1) (right) spectra towards complex \#2. The scale
for each box is $-30\le v_{hel}\le60$\,\kms\ and $-50\le T_{mb}\le110$\,mK.}
\label{fig_cloud2}
\end{figure*}

Towards complex \#2 we find significant variation of the line ratio of
the lower two CO transitions. Spectra towards the center of that complex
are displayed in Fig. \ref{fig_cloud2}. The center position is easily
detected in both the (1$\to$0) and in the (2$\to$1) transition,
whereas in the surrounding positions the (2$\to$1) transition is
hardly detectable towards single positions. However, (2$\to$1) emission is
clearly present in the averaged spectrum. Values derived from a
Gaussian analysis of the spectra for the center position and the
average of the eight surrounding positions are listed in
Table \ref{tab_cloud2}. The line ratio is $R_{21}=2.0\pm0.4$ for the
center position and $R_{21}=0.7\pm0.2$ for the surrounding area. In
Sect. \ref{excitation} we discuss different beam filling and variation
of the excitation conditions as possible causes for the variation of
the line ratio.

\begin{table}
\caption{Line parameters of CO spectra of complex \#2}
\begin{tabular}{l l l l l}
\noalign{\hrule}
\noalign{\smallskip}
Transition & $T_{mb}$ & $rms$  & $v_{hel}$ & $\Delta v$ \\
           & (mK)      & (mK)    & (\kms)    &   (\kms) \\
\noalign{\hrule\smallskip\hskip 1cm Center Position}
\noalign{\smallskip\hrule\smallskip}
$(1\to0)$ & 63 & 8 & $19.2\pm0.4$ & $8.7\pm0.9$ \\
$(2\to1)$ & 102 & 18 & $20.0\pm0.7$ & $10.7\pm1.5$ \\
\noalign{\hrule\smallskip\hskip 1cm Average of surrounding Positions}
\noalign{\smallskip\hrule\smallskip}
$(1\to0)$ & 18 & 3 & $16.6\pm0.6$ & $16.7\pm1.4$ \\
$(2\to1)$ & 11 & 4 & $13.9\pm2.2$ & $19.7\pm4.6$ \\
\noalign{\hrule\smallskip}
\label{tab_cloud2}
\end{tabular}
\end{table}

\section{Physical parameters of the molecular gas \label{physpar}}

The molecular complex near NGC\,3077 discussed here is larger than
many complexes in other galaxies; e.g., the complex within NGC\,3077
itself has a size (FWHM) of only 320\,pc (Becker et al.\
\cite{becker:etal89}). The complexes  near
NGC\,3077 are well confined over a narrow range in velocity.
  Cohen et al.\ (\cite{cohen:etal88}) report a full
extent of the 30 Dor complex in the LMC of 2400\,pc. 
Most Galactic molecular clouds have sizes below 60\,pc (Solomon et
al. \cite{solomon:etal87}). The Orion A \& B complexes (see Dame et
al. \cite{dame:etal87}) placed at the distance of NGC\,3077 would be
detectable with the sensitivity of our observations, however just in
one single spectrum.  {\it The complex near NGC\,3077 thus belongs to
one of the largest concentrations of molecular gas in the local
universe.}

\subsection{Excitation conditions of the molecular gas \label{excitation}}

We will now use the ratios of the integrated line intensities to get a
handle on the physical parameters. The most stringent limitations on
the kinetic temperature and H$_2$ volume density can be derived for
complex \#1 for which we have observed 3 rotational transitions. In the
following we apply a simple large-velocity gradient code (LVG) which
adopts constant kinetic temperature and H$_2$ volume density
throughout the cloud (see e.g. de Jong et al. \cite{dejong:etal75} for
details).  We calculate the line temperatures for monotonous velocity
gradients of $v_{\rm grad}=1$\kms\,pc$^{-1}$ and $v_{\rm
grad}=10$\kms\,pc$^{-1}$, and for CO abundances of
[CO/H$_2]=8\times10^{-5}$ and [CO/H$_2]=1\times10^{-5}$.  The higher
CO abundance is a representative value for Milky Way clouds
(e.g. Blake et al.\ \cite{blake:etal87}), the other was chosen to
study the effects of lower metallicity.

For complex \#1 the observed line ratios can only be matched if the
kinetic temperature is about 10\,K. The H$_2$ volume density should be
between 600\,cm$^{-3}$, in the case of high abundance and low velocity
gradient, and about 10000\,cm$^{-3}$, in the case of low abundance and
high velocity gradient. The kinetic temperature is the similarly low
as that for cold dark clouds in the Milky Way with no embedded
massive stars, but significantly lower than that of clouds with
on-going massive star-formation (e.g. Jijina et al. \cite{jijina:etal99}).
Note that this is a large-scale ($\sim 1$\,kpc) average value for the
complex which does not exclude localized small regions which may be
warmer and thus may be heated by low--level star--formation.

For complex \#2 we observed only two transitions.  The observed line
ratio can be explained within a wide range of parameters.  We can
however argue that because the ratio for the outer area of that cloud
is similar to that of complex \#1 as a whole that also the physical
conditions may be similar, i.e. low kinetic temperature and medium to
high volume density. In the center of complex \#2 the ratio is much
higher. Part of this might be due to the different beam areas for the
two transitions, 22$''$ for the (1$\to$0) transition and 11$''$ for
the (2$\to1$) transition. This difference can affect the ratio by a
factor of up to 4, if the observed cloud is a point source for both
telescope beams.

We argue, however, that this effect only partly can account for the
variation within complex \#2, because it is extended in both
transitions (as indicated by the presence of (2$\to1$) emission in the
averaged spectra surrounding the centre); in addition the line width
is too wide in the (2$\to1$) transition to be due to one single small
molecular cloud. It is therefore more likely that indeed the kinetic
temperature and/or the H$_2$ volume density rise in complex
\#2. This could be attributed to suggested on--going low--level
star formation in the `Garland' region near complex \#2 (Karachentsev et
al.\ \cite{karachentsev:etal85b}).

\subsection{Molecular cloud masses \label{comass}}

One important yet difficult to determine physical parameter is the
mass of a molecular complex. Determination of the mass is usually
based on either the assumption of virialization and/or application of
a $X_{\rm CO}=N({\rm H}_2)/W_{\rm CO}$ conversion factor. In this
section we apply both methods and discuss their pros and cons.

To estimate the $X_{\rm CO}$ factor we compare the CO luminosity for a
given line width with that of a cloud with the same line width but
with known molecular mass (e.g. Cohen et al.\
\cite{cohen:etal88}). Using Fig.~2 of Cohen et al.\ we find that complex
\#1 lies within the range of values spanned by Galactic clouds, whereas
complex \#2 lies in the range spanned by LMC clouds. This indicates that
the tidal arm clouds have CO luminosities for a given line width in
between those of Milky Way and LMC clouds. The $X_{\rm CO}$ factor of
the Milky Way ($\sim2.5\times10^{20}$\,cm$^{-2}$\,(K\,\kms)$^{-1}$) is a
factor of 6.7 lower than that of the LMC
($1.7\times10^{21}$\,cm$^{-2}$\,(K\,\kms)$^{-1}$, Cohen et al.\
\cite{cohen:etal88}).  Thus we use an $X_{\rm CO}$ value for the tidal
arm clouds which is in between both values, $X_{\rm
CO}=8\times10^{20}$\,cm$^{-2}$\,(K\,\kms)$^{-1}$.  This leads to
molecular masses for complex \#1 of $M_{X,1}=1.7\times10^7$\,\mo\ and
for complex \#2 of $M_{X,2}=0.4\times10^7$\,\mo, (corrected for the
contribution of He).

If we adopt a $1/r$ density profile through the clouds the assumption
of virialization leads to masses for complex \#1 of $M_{\rm
vir,1}=2.4\times10^7$\,\mo\ and for complex \#2 of $M_{\rm
vir,2}=1.6\times10^7$\,\mo.  Given the uncertainties in both methods
the masses for complex \#1 agree well, however those for complex \#2
are discrepant by a factor of 4.  At this point we can only speculate
which method gives the better estimate for the true molecular masses.

\subsection{The relation to the \hi\ gas \label{HI}}

\begin{figure*}
 \resizebox{17.5cm}{!}{\includegraphics{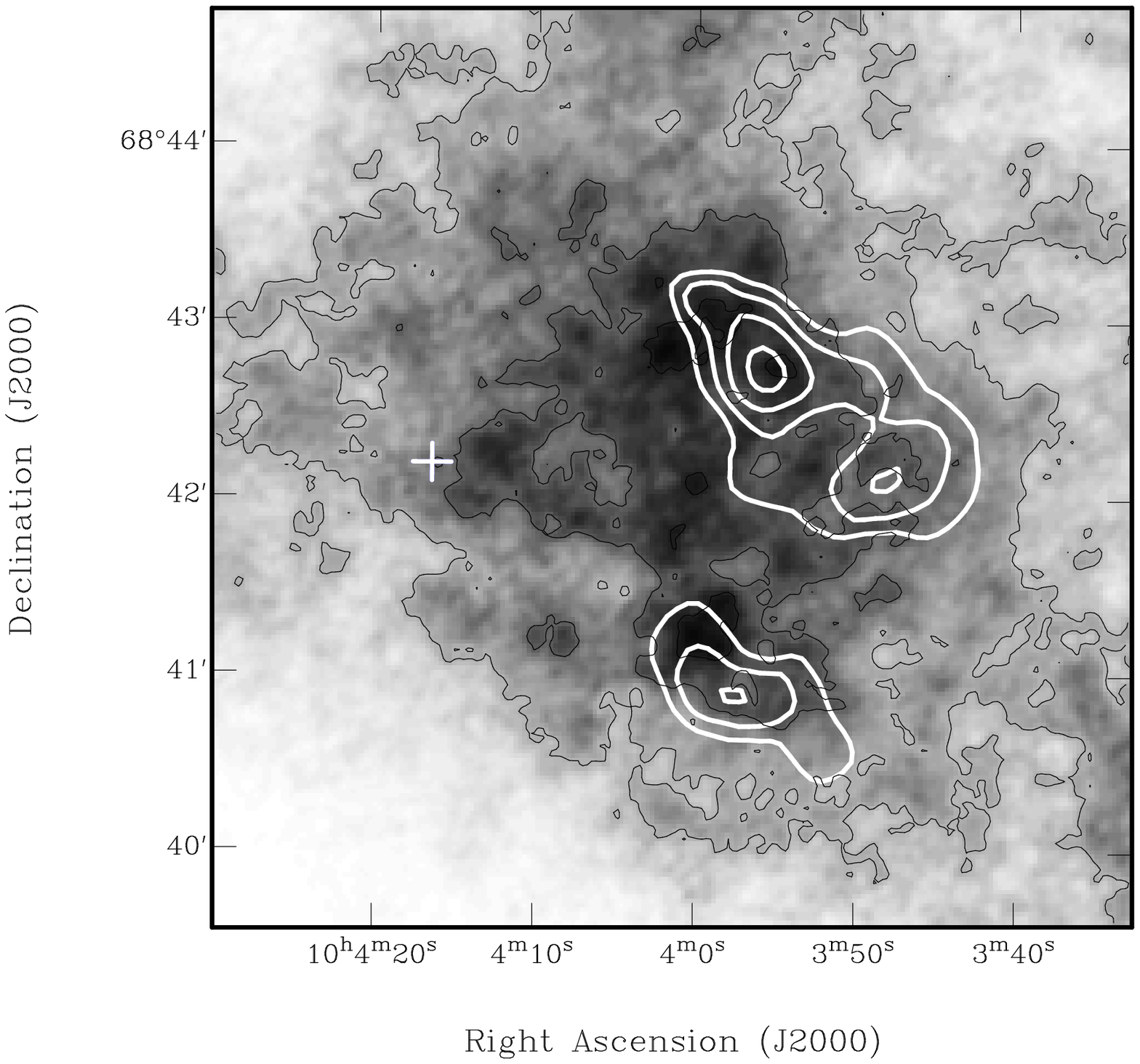}}
\caption{Overlay of 
CO ($J=1\to0$) contours on an integrated \hi\ map (grey scale).
Thick white (CO) contours are every 0.14\,K\,\kms\ starting at 0.28\,K\kms.
Thin black contours represent \hi\ column densities of 1$\times$, 1.5$\times$,
 and 2$\times10^{21}$\,cm$^{-2}$. The + marks the position of complex \#3.
Note that the area has not fully been searched for CO, s. Fig. \ref{comap}
for the sampling.}
\label{hi-co}
\end{figure*}

Fig. \ref{hi-co} shows a comparison of the distribution of the CO gas with
that of atomic hydrogen. We use our high-angular resolution (13$''$) 21\,cm map
as obtained with the VLA (Walter \& Heithausen \cite{walter:heithausen99}). It
is obvious, that there is no direct
correlation between the intensities of \hi\ and
CO. CO is not concentrated at the peak of the \hi\ column density, but
rather found on the outer area of the $1.5\times10^{21}$\,cm$^{-2}$ contour.
The average \hi\ column density associated with complex \#1 and 2 is
$1.6\times10^{21}$\,cm$^{-2}$, that for complex \#3 is
$1.4\times10^{21}$\,cm$^{-2}$. At position \#4 (Table \ref{coclouds}), where
 no CO was found, the associated \hi\ column density is
$1.1\times10^{21}$\,cm$^{-2}$.

The threshold where to find molecular gas depends on both metallicity
and radiation field (e.g. Pak et al. \cite{pak:etal98}). The shielding
\hi\ column density in the tidal arms around NGC\,3077 is slightly
higher than values in other galaxies. In the Milky Way the H$_2$
threshold appears at $0.6\times10^{21}$\,cm$^{-2}$ (Savage et
al. \cite{savage:etal77}). CO observations of M\,31 suggest a
threshold of about 10$^{21}$\,cm$^{-2}$ (Lada et
al. \cite{lada:etal88}). Young \& Lo (\cite{young:lo97}) found a
threshold of $0.1-0.2\times10^{21}$\,cm$^{-2}$ for the dwarf
elliptical galaxies NGC\,185 and NGC\,205; they attribute the
difference to the Milky Way value and M\,31 to the lower radiation
field in these dwarf ellipticals. We only can speculate what causes
the higher threshold in our complex. The radiation field is probably
low because we do not see strong associated star forming regions. 

The total \hi\ mass of the tidal arm feature around NGC\,3077 was
found to be $M$(\hi)=$(3-5)\times10^8$\,\mo\ (van der Hulst
\cite{vanderhulst79}, Walter \& Heithausen
\cite{walter:heithausen99}), depending on integration boundaries. If we
regard only the areas where CO is detected we find an \hi\ mass for
complex \#1 of about $2.7\times10^7$\,\mo\ and for complex \#2 of about
$0.8\times10^7$\,\mo. These atomic masses are comparable to the
molecular masses derived by our adopted $X_{\rm CO}$ factor (see
Sect.~\ref{comass}).

\section{Origin and future of the molecular complex \label{origin}}

Evidently, the galaxies NGC\,3077, M\,81, and M\,82 have undergone
some violent interaction in the past. Visible signs of this
interaction are the long tidal arms seen in the 21\,cm line of atomic
hydrogen (van der Hulst \cite{vanderhulst79}, Yun et al.\
\cite{yun:etal94}) connecting the galaxies.  According to numerical
simulations the closest encounter between NGC\,3077 and M\,81, which
presumably redistributed much of the interstellar gas in the outskirts
of NGC\,3077, happened some $3\times10^8$\,years ago (Yun et
al. \cite{yun:etal93}).

Now, as already discussed in Walter \& Heithausen
(\cite{walter:heithausen99}) it should be stressed that it is very
surprising that a huge amount of molecular gas is present at all in
the tidal arm. Not only are molecular gas and hence metals present far
off a galaxy but the total mass of the molecular gas in the tidal arm
($2-4\times10^7$\,M$_{\odot}$) is possibly higher than the entire
molecular mass within NGC\,3077 itself
($\sim1\times10^7$\,M$_{\odot}$, Becker et al.\
\cite{becker:etal89}). It is not clear if the same ratio also
holds for the total metal content in the tidal arm and in
NGC\,3077. In any case our finding implies that huge amounts of
metal enriched material can be removed from a galaxy due to tidal
interactions. This has not only important consequences for the
chemical history of a single galaxy which might have undergone some
interaction -- interactions also seem to be able to enrich the
intergalactic medium.

Regarding the molecular complex, an important question is
which process condenses the
molecular gas. Are the complexes pressure confined or gravitationally
bound?  While it is hard to imagine that a structure of kpc size is
pressure confined, the intergalactic pressure is most likely too
low. If we use the results from our excitation study
(Sect. \ref{excitation}) the internal pressure in the molecular gas is
$n\times T\ge10^4$\,cm$^{-3}$\,K. From X-ray observations with ROSAT
Bi et al. (\cite{bi:etal94}) found an intergalactic volume density of
less than $1.5\times10^{-3}$\,cm$^{-3}$ in the region of the tidal arm
around NGC\,3077. The intergalactic gas thus must have a temperature
of more than $10^7$\,K to confine the molecular gas, which is 
unlikely.  With an adopted temperature of $1-2\times 10^6$\,K, 
similar to that of the galactic corona 
(Wolfire et al. \cite{wolfire:etal95}),
the intergalactic pressure is less than $\approx2000$\,K\,cm$^{-3}$, 
too low to confine the complexes.

We thus conclude that the molecular clouds are indeed gravitationally
stable objects. The low observed main beam brightness temperature for
all clouds indicates a low beam filling. If complex \#1 is as cold as
derived from our excitation study (Sect. \ref{excitation}) the observed
mainbeam brightness temperature of only about 70\,mK translates into a
beam filling of $1/100$.  This implies that the large complex will
probably break up into several smaller clouds when observed at higher
angular resolution. Because the complex is extended the arguments hold
for all of the observed positions. This implies that the single clouds
are spread over the area with an equivalent radius of 700\,pc, and are
not confined to one single massive cloud. High angular resolution
interferometric CO studies will allow to investigate this situation further.

\section{Conclusions \label{conclusions}}

Our new observations have revealed extensive molecular gas in the
tidal arms near NGC\,3077. The CO emission is much more extended than
previously thought -- the detected complexes range among the most
extended complexes in the local universe. We have detected at least
three independent complexes of molecular gas. The complexes are most
probably gravitationally bound  and have formed {\it in
situ}. For the largest of the complexes our multi-transition CO study
shows that the gas must be cold ($\approx$10\,K), thus massive star
formation is not going on at a significant level. 
 Whether or not the chain of blue stars
(the `Garland') found in this region (Karachentsev et al.\
\cite{karachentsev:etal85a}, \cite{karachentsev:etal85b}) is associated 
with the molecular complex is an open question which is currently
under investigation by us.

The fact that CO is found in tidal arms implies that galaxy
interactions can efficiently remove enriched material from a galaxy's
body hence influencing it's chemical history. This also has important
implications for the chemical enrichment of the intergalactic medium
(IGM), especially at larger look--back times in the universe where
galaxy interactions may have been more frequent.

Since the tidal arm with its newly discovered molecular complexes has
all the ingredients to form
stars in the future (i.e., atomic and molecular gas), our new
observation confirm our previous speculation (Walter \& Heithausen
\cite{walter:heithausen99}) that we are witnessing the birth of a
dwarf galaxy where star formation might start in the near future. We
are therefore in the fortunate situation to witness a process which
may have created a substantial number of today's dwarf galaxies in the
past.

\begin{acknowledgement}
We thank Christian Henkel for critically reading the manuscript and Andrea 
Tarchi for help with part of the observations.
FW acknowledges NSF grant AST 96--13717. The KOSMA 3\,m telescope is
operated by the Uni\-versity of Cologne, Germany, and supported by the
DFG grant SFB 301, as well as special funding from the Land NRW,
Germany. The observatory is administered by the Internationale
Stif\-tung Hochalpine Forschungsstationen Jungfraujoch und Gornergrat,
Bern, Switzerland.
\end{acknowledgement}

{}

\end{document}